\documentclass{iaus}
\usepackage{graphicx}

\title[Synchrotron Outbursts in Galactic and Extra-galactic Jets, Any Difference?] 
{Synchrotron Outbursts in Galactic and Extra-galactic Jets, Any Difference?}

\author[M. T\"urler \& E.~J. Lindfors]   
{M. T\"urler$^{1,2}$ \and E.~J. Lindfors$^{3,4}$}

\affiliation{$^1$INTEGRAL Science Data Centre, ch. d'Ecogia 16, 1290 Versoix, Switzerland\break email: marc.turler@obs.unige.ch\\[\affilskip]
$^2$Geneva Observatory, University of Geneva, ch. des Maillettes 51, 1290 Sauverny, Switzerland\\[\affilskip]
$^3$Tuorla Observatory, V\"ais\"al\"a Institute of Space Physics and Astronomy, University of Turku, 21500 Piikki\"o, Finland. email: elilin@utu.fi\\[\affilskip]
$^4$Mets\"ahovi Radio Observatory, Helsinki University of Technology, 02540 Kylm\"al\"a, Finland}
\pubyear{2006}
\volume{238}  
\pagerange{001--999}
\date{??? and in revised form ???}
\jname{Black Holes: from Stars to Galaxies -- across the Range of Masses}
\editors{V. Karas \& G. Matt, eds.}
\begin{document}

\maketitle

\begin{abstract}
We discuss differences and similarities between jets powered by super-massive
black holes in quasars and by stellar-mass black holes in microquasars. The
comparison is based on multi-wavelength radio-to-infrared observations of the
two active galactic nuclei 3C 273 and 3C 279, as well as the two galactic
binaries GRS 1915+105 and Cyg X-3. The physical properties of the jet are
derived by fitting the parameters of a shock-in-jet model simultaneously to all
available observations. We show that the variable jet emission of galactic
sources is, at least during some epochs, very similar to that of extra-galactic
jets. As for quasars, their observed variability pattern can be well reproduced
by the emission of a series of self-similar shock waves propagating down the jet
and producing synchrotron outbursts. This suggests that the physical properties
of relativistic jets is independent of the mass of the black hole.
\end{abstract}

\firstsection 
\section{Introduction}
Quasars and microquasars - their galactic analogs - are relativistic jet sources
thought to be powered by super-massive and stellar-mass black holes,
respectively (see Mirabel \& Rodr\'{\i}guez \cite{MR98} for a review). Although the radio
emission of microquasars tends to show a greater variety of behaviours, during
some activity periods their radio variability pattern resembles that of
quasars, but on much shorter timescales (hours or days instead of years). It is
now rather well established for quasars that outbursts identified in their radio
lightcurves are related to moving structures in the jet imaged with radio
interferometric techniques. There is also growing evidence that these structures
emitting synchrotron radiation are propagating shock waves along the jet flow
rather than discrete ejected plasma clouds (Kaiser \etal\ \cite{KSS00}; T\"urler \etal\ \cite{TCC04}).

Based on these facts, a natural step forward is to confront theoretical
predictions of shock wave models with the observed evolution of the outbursts.
The approach we developed since several years is to try to fit a series of
self-similar model outbursts to the multi-wavelength monitoring observations of
some of the best studied objects. Until now, we studied data sets of four
objects: the quasar 3C~273 (T\"urler \etal\ \cite{TCP99}; \cite{TCP00}), the
blazar 3C~279 (Lindfors \etal\ \cite{LTV06}) and the microquasars GRS~1915+105
(T\"urler \etal\ \cite{TCC04}) and Cyg~X-3 (Lindfors \etal\ in prep.). In this
short contribution, we try to compare and discuss the jet properties derived for
these four sources as one of the first attempts to sketch out a more global
picture of black hole jets across the range of mass.

\section{Data and Method}

\begin{figure}
  \begin{center}
  \includegraphics[width=0.8\textwidth]{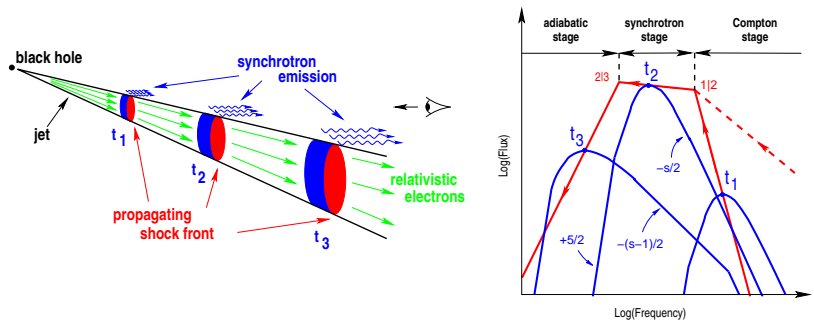}
  \caption{Schematic view of a propagating shock wave in a relativistic jet and
  the three-stage evolution of the associated synchrotron outburst according to
  the model of Marscher \& Gear (\cite{MG85}) and with the modification proposed
  by Bj\"ornsson \& Aslaksen (\cite{BA00}) (dashed line)}\label{fig:1}
  \end{center}
\end{figure}

A description of the data sets and the method can be found in our previous
publications listed above. We just recall here that for 3C~279 and 3C~273 we use
data spanning one and two decades, respectively, whereas for GRS 1915+105 we use
the data of 15 May 1997 in the plateau/state C/class $\chi$ state (Mirabel \&
Rodr\'{\i}guez \cite{MDC98}) and for Cyg~X-3 the outbursts of Feb.--Mar. 1994
(Fender \etal\ \cite{FBW97}). The main change with respect to previous studies
is that we use here consistently for all sources the modification proposed by
Bj\"ornsson \& Aslaksen (\cite{BA00}) for the initial Compton stage of the
Marscher \& Gear (\cite{MG85}) shock model in the case of only first-order
Compton cooling. This flattens the initial rise of the spectral
turnover with decreasing frequency, as shown schematically in
Fig.~1. In the case of 3C~273 the jet parameters are such that this modification
made the Compton stage completely flat or even inverted (decreasing turnover
flux with time), which would make the model incompatible with infrared data.
This problem could however be solved by assuming that the shocked material is
viewed sideways (Marscher \etal\ \cite{MGT92}). Such an assumption is not
unreasonable in sources with highly superluminal jets, as the jet angle $\theta$
to the line of sight for this to happen is defined by $\cos(\theta)=\beta
\Leftrightarrow\ \sin(\theta)=1/\gamma$, which is the same condition that
maximizes the apparent transverse velocity $\beta_{\mathrm{app}}$ for a given
bulk velocity $\beta=v/c$. We therefore use here a sideways orientation of the
jet for both extra-galactic sources and a face-on orientation for both
microquasars.

\section{Results and Discussion}

\begin{table}\def~{\hphantom{0}}
  \begin{center}
  \caption{Physical jet properties of the four objects derived for the average model outburst at the time when it reaches a maximum flux (i.e. around the transition to the final decay stage).}
  \label{tab:1}
  \begin{tabular}{@{}lccccccccc@{}}\hline
     Object		& $M_{\mathrm{BH}}$ & $\Delta\,t_{\mathrm{obs}}$ & $\Delta\,l$ & $\theta_{\mathrm{src}}$ & $B$ & $K$ & $s$ & $u_{\mathrm{e}}$ & $u_{B}/u_{\mathrm{e}}$ \\
			& [ M$_{\odot}$ ] & [ s ]	& [ AU ]	& [ mas ] & [ G ] & [ erg$^{s-1}$/cm$^3$ ] &  & [ erg/cm$^3$ ] &  \\\hline
     3C 273 $^1$	& 6.6\,10$^{9}$	& 3.1\,10$^{7}$	& 3.1\,10$^{6}$	& 0.39	& 0.13	& 6.7\,10$^{-7}$& 2.52	& 1.3\,10$^{-4}$& 5.6		\\
     3C 279 $^2$	& 3\,10$^{8}$ 	& 3.1\,10$^{7}$	& 4.7\,10$^{6}$	& 0.25	& 0.26	& 8.5\,10$^{-7}$& 2.09	& 1.4\,10$^{-5}$& 195		\\
     Cyg X-3 $^3$	& $<$\,3.6 	& 8.6\,10$^{4}$	& 6.6\,10$^{2}$	& 4.6	& 0.83	& 1.9\,10$^{-4}$& 1.68	& 7.9\,10$^{-5}$& 344		\\
     GRS 1915 $^4$	& 14 		& 1.8\,10$^{3}$	& 3.1		& 0.02	& 0.15	& 6.7\,10$^{2}$	& 1.79	& 1.4\,10$^{3}$	& 6.3\,10$^{-7}$\\
     GRS 1915 $^5$	& 14 		& 1.8\,10$^{3}$	& 6.7		& 0.07	& 7.5	& 3.4\,10$^{-1}$& 1.79	& 4.7\,10$^{-1}$& 4.8		\\\hline
  \end{tabular}
 \end{center}
  $^1$ $M_{\mathrm{BH}}$ from Paltani \& T\"urler (\cite{PT05}) and with $\beta_{\mathrm{app}}$\,=\,10 \& $\theta$\,=\,10$^{\circ}$ (Savolainen \etal\ \cite{SWV06})\\
  $^2$ $M_{\mathrm{BH}}$ from Wang \etal\ (\cite{WLH04}) and with $\beta_{\mathrm{app}}$\,=\,10 (Jorstad \etal\ \cite{JML04}) \& $\theta$\,=\,5$^{\circ}$ (see text)\\
  $^3$ $M_{\mathrm{BH}}$ from Stark \& Saia (\cite{SS03}) and $\beta$\,=\,0.81, $\theta$\,=\,14$^{\circ}$ \& $d$\,=\,10\,kpc (Mioduszewski \etal\ \cite{MRH01})\\
  $^4$ $M_{\mathrm{BH}}$ from Greiner \etal\ (\cite{GCM01}) and $\beta$\,=\,0.6, $\theta$\,=\,61$^{\circ}$ \& $d$\,=\,9\,kpc (T\"urler \etal\ \cite{TCC04})\\
  $^5$ alternative with $\beta$\,=\,0.9, $\theta$\,=\,55$^{\circ}$ \& $d$\,=\,7\,kpc (uncertainty on $d$, see Fender \etal\ \cite{F03})\\
\end{table}

Due to lack of space we cannot present here the derived evolution of the average
outburst in the four sources. The main differences we obtain with the changes to
the model described above, is that in general we do not obtain an almost flat synchrotron stage, but one with a much steeper increase of flux with turnover frequency, which becomes more difficult to distinguish from the now shallower Compton stage. There is therefore some uncertainty whether the synchrotron stage
exists at all or whether there is a direct transition from the Compton to the adiabatic stage. Another interesting point is that for the initial phases of the outburst evolution we suspect the spectral turnover $\nu_{\mathrm{m}}$  not to be due to synchrotron self-absorption as assumed, but to a low energy cut-off of the electron energy distribution. Evidence for this is the shallow spectral slope we find below the turnover frequency and which we describe in our modelling by an inhomogeneous synchrotron source. However, one would rather expect the source to become more inhomogeneous with time as the source increases in size, whereas we observe the opposite behaviour. An alternative explanation is that this flatter spectral index mimics the characteristic $\nu^{1/3}$ slope expected when self-absorption becomes important at a lower frequency than the synchrotron frequency associated to the electrons with minimal energy (see e.g. the spectral shape in the upper panel of Fig.~1 of Granot \& Sari \cite{GS02}). This interpretation could also explain the too rapid evolution of the turnover frequency with time that we observe in both galactic sources. Currently, this is described in our modelling by a non-conical jet with an opening angle increasing with time, but the physical justification for such a trumpet-like shape is not clear.

More quantitatively, we made a first attempt to derive the physical properties
of the jets, which we summarize in Table~1. These values have been derived for
the particular point when the synchrotron spectrum of the average outburst
reaches its maximum flux, which happens here for all sources at the transition
to the final adiabatic cooling phase. A critical parameter is the angular size
$\theta_{\mathrm{src}}$ of the source of synchrotron emission, which we assume
to be equal to the width of the jet. By assuming a same jet opening half-angle
of 2$^{\circ}$ for all sources, $\theta_{\mathrm{src}}$ can thus be calculated
from the length along the jet $\Delta\,l$ traveled by the shock during the
observed time interval $\Delta\,t_{\mathrm{obs}}$ needed for the outburst to
reach a maximal flux. This length itself depends on good estimates of the
object's distance $d$, the jet angle $\theta$ to the line of sight, and the
apparent $\beta_{\mathrm{app}}=v_{\mathrm{app}}/c$ or real $\beta=v/c$ jet
speed. We can then simply use Eqs.~(3) to (5) of Marscher (\cite{M87}) to
calculate the magnetic field strength $B$, the normalization $K$ of the electron
energy distribution $N(E)=K\,E^{-s}$ and the energy density $u_{\mathrm{e}}$ of
relativistic electrons. For the latter we use a ratio $\nu_2/\nu_1$ of 10$^4$
between the highest $\nu_2$ and the lowest $\nu_1$ frequency considered for
integration. Although these formula apply strictly to a homogeneous and
spherical synchrotron source, we do not expect the emitting material in the jet
to depart a lot from this ideal case at the start of the final decay stage.
Finally, we calculate the energy density of the magnetic field
$u_{B}=B^2/(8\pi)$ and the ratio $u_{B}/u_{\mathrm{e}}$, which is extremely
sensitive to the source size as: $u_{B}/u_{\mathrm{e}} \propto
\theta_{\mathrm{src}}^{~17}$. It is thus quite surprising to get values close to
equipartition (i.e. $u_{B}\approx u_{\mathrm{e}}$) for all objects when assuming
a same jet opening half-angle of 2$^{\circ}$ and realistic values for the jet
orientation and speed as given in the footnotes of Table~1. For GRS~1915+105
there is an important uncertainty on the distance $d$ and jet velocity $\beta$
(Fender \etal\ \cite{F03}) and actually we find that higher jet speeds and
smaller distances favor equipartition. For 3C~279 we note that taking a viewing
angle $\theta$ of 1$^{\circ}$ (Jorstad \etal\ \cite{JML04}) instead of
5$^{\circ}$ results in extreme values of $B=4.6\,10^2$\,G and
$u_{\mathrm{e}}=3.8\,10^{-14}$\,erg\,cm$^{-3}$ leading to a ratio
$u_{B}/u_{\mathrm{e}}$ of $2\,10^{17}$, which seems very unrealistic.

\section{Conclusion}
The observed properties of synchrotron outbursts in microquasars appears to be
quite similar, at least during some period of activity, with the behaviour of
quasars. We find that timescales and physical sizes of the jet do scale with the
black hole mass, but not strictly linearly as expected (e.g. Mirabel \&
Rodr\'{\i}guez \cite{MR98}). For the observed timescales we find very roughly a
square root dependence $\Delta\,t_{\mathrm{obs}}\propto M_{\mathrm{BH}}^{1/2}$,
becoming more linear for the intrinsic lengthscale $\Delta\,l$ of the jet, which
is corrected for orientation and relativistic effects. Appart from this scaling
the physical properties of the jets, like the magnetic field and the electron
energy density, are found to be very similar in all sources except GRS~1915+105,
which has significantly higher values. The only clear difference we find between
galactic and extra-galactic jets is a harder electron energy distribution in
microquasars, with an index $s$ being clearly below 2 in Cyg~X-3 and
GRS~1915+105.

Our method of fitting multi-wavelength lightcurves with model outbursts is
getting to the point where we can test different models and constrain the
physics of relativistic jets. There are indications that the standard shock
model of Marscher \& Gear (\cite{MG85}) has to be modified to describe the
observed evolution of synchrotron outbursts. Apart from the modifications proposed by Bj\"ornsson \& Aslaksen (\cite{BA00}), we find evidence
that the spectral turnover might rather be due to a low energy cut-off of the
electron energy distribution than to synchrotron self-absorption during the initial phases of the outburst.

More information, figures and animations at:
http://isdc.unige.ch/$\sim$turler/jets/

\end{document}